\documentclass[12pt]{article}%
\begin{document}%
\markright{Interpreting doubly special relativity\hfil}%
\def\Universita{Universit\`a}%
\title{\bf \LARGE Interpreting doubly special relativity as a
modified theory of measurement}%
\author{Stefano Liberati~$^*$, Sebastiano Sonego~$^\dagger$, and
Matt Visser~$^\ddagger$\\[2mm]%
{\small\it
\thanks{\tt liberati@sissa.it; http://www.sissa.it/\~{}liberati}%
\ International School for Advanced Studies}\\%
{\small \it Via Beirut 2-4, 34013 Trieste, Italy}
\\ {\small \it%
and INFN, Trieste, Italy}%
\\[4mm]%
{\small\it\thanks{\tt sebastiano.sonego@uniud.it}%
\ {\Universita} di Udine, Via delle Scienze 208, 33100 Udine, Italy}\\[4mm]%
{\small\it
\thanks{\tt matt.visser@mcs.vuw.ac.nz; http://www.mcs.vuw.ac.nz/\~{}visser}%
\ School of Mathematics, Statistics, and Computer Science}\\%
{\small \it Victoria University of Wellington}\\%
{\small \it PO Box 600, Wellington, New Zealand}}%
\date{{\small 22 October 2004; {\LaTeX-ed \today}; gr-qc/0410113}}%
\maketitle%
\begin{abstract}%
In this article we develop a physical interpretation for the deformed
(doubly) special relativity theories (DSRs), based on a modification
of the theory of measurement in special relativity.  We suggest that
it is useful to regard the DSRs as reflecting the manner in which
quantum gravity effects induce Planck-suppressed distortions in the
measurement of the ``true'' energy and momentum. This interpretation
provides a framework for the DSRs that is manifestly consistent,
non-trivial, and in principle falsifiable.  However, it does so at the
cost of demoting such theories from the level of ``fundamental''
physics to the level of phenomenological models --- models that should
in principle be derivable from whatever theory of quantum gravity one
ultimately chooses to adopt.%
\vspace*{5mm}%
\\%
\end{abstract}%
\def\g{{\mbox{\sl g}}}%
\def\Box{\nabla^2}%
\def\d{{\mathrm d}}%
\def\i{{\mathrm i}}%
\def\ie{{\em i.e.\/}}%
\def\eg{{\em e.g.\/}}%
\def\etc{{\em etc.\/}}%
\def\etal{{\em et al.\/}}%
\clearpage%

\section{Introduction}%
\label{sec:introduction}%
\setcounter{equation}{0}%

The search for observable effects of quantum gravity has been one of
the driving trends in recent years. Several results in the context of
string theory~\cite{KS}, loop quantum gravity~\cite{loopqg} and other
candidate models~\cite{Burgess} for quantum gravity, have led
researchers to focus mostly on modifications of the dispersion
relations for elementary particles, leading to deviations from
standard Lorentz invariance.  Generically these modified dispersion
relations can be cast in the form\footnote{We work in units with
  $c=1$.}
\begin{equation}%
E^2=p^2+m^2+f(E,p;\kappa)\;,%
\label{eq:disprel}%
\end{equation}%
where $\kappa$ denotes the mass scale at which the quantum gravity
corrections become appreciable.  Normally, one assumes that $\kappa$
is of order the Planck mass: $\kappa \sim M_{\rm P} \approx 1.22\times
10^{19}\;$GeV.  Most interestingly, it was shown that several
significant constraints can be put on the intensity of the Lorentz
violating term $f(E,p;\kappa)$ using current experiments and
observations~\cite{QGph}.%

An open issue is the interpretation of the origin of such deformed
dispersion relations.  There is an extensive literature in which an
explicit breakdown of Lorentz invariance has been considered (see
\cite{breaking} and references therein).  However, some authors have
tried to see if a more conservative generalization of the Lorentz
transformations could be found, in order to save the equivalence of
inertial frames in special relativity~\cite{AC1,ms,AC3}.  In
particular the DSRs (deformed or doubly special relativity theories)
attempt to ``deform'' special relativity in momentum space, by
introducing non-standard ``Lorentz transformations'' that leave the
modified dispersion relations above invariant.%

Unfortunately, Lorentz invariance provides an extremely strong and
rigid framework for particle physics, and while it is relatively easy
to ``break'' Lorentz invariance, it is much more difficult to
``deform'' it without ``breaking'' it.  This has led to considerable
debate concerning the physical status of DSRs, with a strong minority
of authors arguing for either the triviality~\cite{Ahlu} or internal
inconsistency~\cite{position,su} of such theories.

In this paper we further investigate the DSR framework and propose an
alternative interpretation that we think is both logically consistent
and non-trivial.  After presenting, in the next section, a very
concise review of the DSR proposal and its open problems we shall
focus, in section \ref{sec:math}, on the momentum space DSR
transformations and their mathematical meaning.  This will lead us to
suggest, in section \ref{sec:measurement}, a physical interpretation
of DSR as a new theory of measurement which could stem from quantum
gravity effects. In section \ref{sec:implications} we shall discuss
how this new framework could be used to solve some of the problems
pointed out by previous authors concerning the DSR proposal.  Finally,
section \ref{sec:conclusions} contains a summary of the main ideas
presented in the paper.%

\section{The DSR framework}%
\label{sec:dsr}%
\setcounter{equation}{0}%

In this section we briefly outline the DSR framework and its open
issues.  This review is in no sense complete as more in-depth
discussions are now available in several published articles (see, \eg,
\cite{GAC-Rio} and references therein).%

\subsection{Deformed Lorentz algebra}%

Consider the Lorentz algebra of the generators of rotations, $L_i$,
and boosts, $B_i$:%
\begin{equation}%
[L_i,L_j] = \i\,\epsilon_{ijk}\; L_k\;; \qquad%
[L_i,B_j] = \i\,\epsilon_{ijk}\; B_k\;; \qquad%
[B_i,B_j] = -\i\,\epsilon_{ijk}\; L_k%
\label{LB}
\end{equation}%
(Latin indices $i,j,\ldots$ run from 1 to 3).  Supplement this with
the following commutators between the Lorentz generators and those of
translations in spacetime (the momentum operators $P_0$ and $P_i$):%
\begin{equation}%
[L_i,P_0]=0\;; \qquad%
[L_i, P_j] = \i\,\epsilon_{ijk}\; P_k\;;%
\label{LP}%
\end{equation}%
\begin{equation}%
[B_i,P_0] = \i\; f_1\left({P\over\kappa}\right) P_i\;;%
\label{BP0}%
\end{equation}%
\begin{equation}%
[B_i,P_j] = \i\left[ \delta_{ij} \; f_2\left({P\over\kappa}\right)
P_0 + f_3\left({P\over\kappa}\right)  {P_i \; P_j\over \kappa}
\right]\;.%
\label{BPj}%
\end{equation}%
Finally, assume%
\begin{equation}%
[P_i,P_j] = 0\;.%
\label{PP}%
\end{equation}%
The commutation relations (\ref{BP0})--(\ref{BPj}) are given in terms
of three unspecified, dimensionless structure functions $f_1$, $f_2$,
and $f_3$, and are sufficiently general to include all known DSR
proposals --- the DSR1~\cite{AC1}, DSR2~\cite{ms}, and
DSR3~\cite{AC3}.  Furthermore, in all the DSRs considered to date, the
dimensionless arguments of these functions are specialized to%
\begin{equation}%
f_i\left({P\over\kappa}\right)  \to f_i\left( {P_0\over\kappa},
{\sum_{i=1}^3 P_i^2\over\kappa^2}\right)\;,%
\end{equation}%
so that rotational symmetry is completely unaffected.  In order that
the $\kappa\to +\infty$ limit reduce to ordinary special relativity we
demand that, in that limit, $f_1$ and $f_2$ tend to $1$, and that
$f_3$ tend to some finite value.%

\subsection{A note on terminology}%

The ``internal'' commutation relations (\ref{LB})--(\ref{LP}), among
the boosts and rotations are not altered in any way --- so the Lorentz
sub-group is not changed at all. This underlies the claim that Lorentz
invariance is not ``broken'' in these theories.  On the other hand,
the DSR group acts on the momenta in a nontrivial manner --- and if we
choose to label ``states'' by the real eigenvalues\footnote{Greek
  indices from the middle of the alphabet, $\mu,\nu,\ldots,$ run from
  0 to 3.} $p_\mu$ of the momentum operators we see that the DSR group
acts non-trivially on states even if it possesses the same number of
symmetry generators as the Lorentz group. This leads to the
nomenclature of a ``deformed'' Lorentz invariance.%

On this terminology we feel that a brief comment is in order. In fact
adopting the above DSR conventions would, if carried to their logical
conclusion, also force one to declare that ``spontaneous symmetry
breaking'' never breaks any symmetry --- simply on the grounds that in
spontaneous symmetry breaking the symmetry group is unaffected, while
it is only the states (and in particular, the vacuum) that then
transform in a nontrivial way.  While DSR does not appear to be an
example of spontaneous symmetry breaking (the uncertainty arising from
the fact that we do not have a precise field theoretic description of
how to implement the DSR algebra), the basic logic is clear: Based on
standard particle physics usage, the Lorentz symmetry in DSR theories
would be classed as ``broken'', and not ``deformed'' (although one
can usefully disagree about whether the breaking is ``soft'',
``hard'', ``spontaneous'', or ``other'').%

\subsection{Nonlinear representations of the Lorentz group}%

In all doubly special relativity theories, there is a claim that the
Lorentz group ``acts nonlinearly on energy and momentum''.  This
amounts to the assertion that physical energy and momentum are
nonlinear functions of a fictitious pseudo-momentum one-form $\pi$,
whose components transform linearly under the action of the Lorentz
group~\cite{jv}. Indeed such behaviour is automatically guaranteed if
the realisation of the Lorentz group on the energy-momentum space is
faithful, \ie, one-to-one \cite{schutz}.  If it were not, then either
(i) the same element of the Lorentz group would act in two different
ways on energy and momentum, or (ii) two different elements of the
Lorentz group would act in the same way.  In case (i) one would need
extra parameters, in addition to those characterising boosts and
rotations, in order to fully specify the transformation.  (The Lorentz
transformations would then be a subgroup of the full physical
transformation group.) The physical meaning of these extra parameters
would be, however, totally obscure. The possibility (ii) conflicts
with the simple experimental fact that, at small energies, different
elements of the Lorentz group are observed to act differently.  Thus,
if $E$ is the ``physical'' energy and $p_i$ are the components of
``physical'' three-momentum, we must have%
\begin{equation}%
p_\mu={\cal F}_\mu(\pi_0,\pi_1,\pi_2,\pi_3;\kappa)\;,%
\label{F}%
\end{equation}%
where $p_0\equiv -E$, and the variables $\pi_\mu$ transform linearly 
under the Lorentz group.  For example, in DSR2, the specific DSR model 
developed by Magueijo and Smolin \cite{ms}:%
\begin{eqnarray}%
&&E=\frac{-\pi_0}{1-\pi_0/\kappa}\;;%
\label{msE}\\%
&&p_i=\frac{\pi_i}{1-\pi_0/\kappa}\;.%
\label{msp}%
\end{eqnarray}%
It is easy to check that while $\pi$ satisfies the usual dispersion
relation $\pi_0^2-\mbox{\boldmath $\pi$}^2=m^2$ (for a particle with
mass $m$), $E$ and $p_i$ satisfy a modified relation%
\begin{equation}%
\left(1-m^2/\kappa^2\right)E^2+2\,\kappa^{-1}\,m^2\,E
-\mbox{\boldmath $p$}^2=m^2\;.%
\label{mod-disp}%
\end{equation}%
In fact, as we said, doubly special relativity has been invented
precisely in order to provide a theoretical background to anomalous
dispersion relations like the one above \cite{AC1}. For a general
theory based on equation (\ref{F}), one can write%
\begin{equation}%
\pi_\mu={\cal G}_\mu(E,\mbox{\boldmath $p$}\,;\kappa)\;,%
\label{G}%
\end{equation}%
with ${\cal G}={\cal F}^{-1}$.  Then, the modified dispersion relation
is%
\begin{equation}%
\eta^{\mu\nu} \; {\cal G}_\mu(E,\mbox{\boldmath $p$}\,;\kappa) \;
{\cal G}_\nu(E,\mbox{\boldmath $p$}\,;\kappa)=-m^2\;,%
\label{mod-disp-gen}%
\end{equation}%
where $\eta_{\mu\nu}$ is the metric tensor of Minkowski spacetime.%

\subsection{Open issues in DSR}%

If DSR is formulated as above --- only in momentum space --- then as
we shall soon see it is an incomplete theory.  Moreover,  since it is 
always possible to introduce the new variables $\pi_\mu$, on which the
Lorentz group acts in a linear manner, the only way that DSR can avoid
triviality is if there is some physical way of distinguishing the
pseudo-energy $\epsilon\equiv -\pi_0$ from the true-energy $E$, and
the pseudo-momentum $\mbox{\boldmath $\pi$}$ from the true-momentum
$\mbox{\boldmath $p$}$ --- otherwise DSR would be no more than a
nonlinear choice of coordinates on momentum space.%

In view of the standard relations $E\leftrightarrow \i\hbar
\partial_t$, $\mbox{\boldmath $p$}\leftrightarrow -\i\hbar
\mbox{\boldmath $\nabla$}$ (which will presumably be modified in some
way in DSR) it is already clear that in order to physically
distinguish the pseudo-energy $\epsilon$ from the true-energy $E$, and
the pseudo-momentum $\mbox{\boldmath $\pi$}$ from the true-momentum
$\mbox{\boldmath $p$}$, one will need to have some idea of how to
relate momenta to position --- at a minimum, one will need to develop
some notion of DSR-spacetime.%

In this endeavour there have been two distinct lines of approach, one
presuming commutative spacetime coordinates, the other trying to
relate the DSR feature in momentum space to a non commutative position
space. In both cases several authors have pointed out major
problems. In the case of commutative spacetime coordinates, some
analyses have led authors to question the triviality~\cite{Ahlu} or
internal consistency~\cite{position,su} of DSR.  On the other hand,
non-commutative proposals \cite{noncommut} are not yet well
understood.%

For these reasons we shall first focus on those problems, or
ambiguities, which are well understood using purely the momentum space
structure of DSR. In particular these include (but this list is not
meant to be exhaustive):%
\begin{itemize}%
\item%
The saturation problem (also known as the ``soccer ball problem''):
How can macroscopic objects, which experimentally certainly can and do
have trans-Planckian total energies, fit into a DSR framework that
typically exhibits a maximum energy of order the Planck energy
\cite{su}?%
\item%
Definition of particle velocities: How are particle velocities to be
defined in DSR? Using phase velocity, group velocity, or something
else \cite{mignemi}?%
\item%
Multiplicity problem: Why are there so many different realizations of
DSRs?%
\end{itemize}%

While the last two problems can be interpreted as ambiguities related
to the incompleteness of the present theory, the first issue
demonstrates a much more serious problem with the multiple particle
sector of the theory. In dealing with collisions or composite objects
it is natural to add linearly the pseudo-momenta $\pi_{\mu}$ and then
transform back to the DSR momenta $p_{\mu}$, so that for $N$ particles
one gets%
\begin{equation}%
p_{\rm tot}={\cal F}\left( \sum_1^N {\cal G}(p_{\mu};\kappa);
\kappa \right)\;.%
\label{eq:comp}%
\end{equation}%
Reduced to the bone, the issue is here related to the fact that the
nonlinear transformation $\cal F$ maps infinity to the Planck
scale (energy or momentum, depending on the particular DSR proposal).
So it would seem that DSR cannot describe objects with energies
(momenta) larger than the Planck scale.  This prediction is 
already very disturbing by itself, but lies also at the origin of 
other unpleasant consequences of DSRs.  For example, the internal 
energy of a gas in the thermodynamical limit $N\to +\infty$ is of 
order $\kappa$.  More generally, one cannot formulate statistical 
mechanics and thermodynamics, because the partition function diverges
\cite{position}.%

To date the various solutions proposed for the saturation problem seem
as problematic as the paradox itself.  For example, it has been
proposed~\cite{ms} to replace (\ref{eq:comp}) with%
\begin{equation}%
p_{\rm tot}={\cal F}_N \left( \sum_1^N {\cal
G}(p_{\mu};\kappa); \kappa \right)\;.%
\label{eq:compN}%
\end{equation}%
As long as we choose ${\cal F}_N$ in such a way that it saturates at
$N \kappa$ instead of $\kappa$, then we can indeed obtain a total
energy that is extensive in the number of particles (so there is at
least a hope of beginning to set up thermodynamics), and a total
energy that can become arbitrarily large (so that we can at least hope
to accurately describe at least the kinematics of planets, stars, and
galaxies).  The canonical choice at this stage is to set ${\cal
  F}_N(\pi;\kappa)={\cal F}(\pi;N\kappa)$.  The crucial point here is
that the resolution of the paradox is obtained at the very high price
of replacing a single DSR algebra with different DSR algebras acting
on each $N$-particle sector of the Fock space.  Alternatively, one
might claim that it is too early to address the problem because of the
lack of a proper field theory (itself due to the lack of a full
comprehension of DSR in coordinate space~\cite{arzano}), but somehow
this is tantamount to ``solving'' a problem with another problem.%

Given the above open issues of DSR, we here wish to restart by looking
at the subject from scratch.  In the following we develop an
interpretation of the DSRs (in terms of a modification of the theory
of measurement in special relativity) which is internally consistent,
mathematically and physically non-trivial, and falsifiable --- three
key tests that any viable physical theory must pass.  Thus by adopting
this interpretation we can guarantee that we are asking (and hopefully
answering) physically meaningful questions.%

\section{The mathematical meaning of $p_\mu$}%
\label{sec:math}%
\setcounter{equation}{0}%

We want to start our investigation from what we know for sure as the
defining properties of all of the DSR theories so far proposed, \ie,
from the relations (\ref{F}). Since the $\pi_\mu$ transform linearly
under the action of the Lorentz group, there is no difficulty in
identifying them as the components of a one-form in Lorentzian
coordinates. But what kind of mathematical objects are the $p_\mu$?
If, by ``action of the Lorentz group'', we simply mean a change of
Lorentzian coordinates, then the $p_\mu$ cannot be scalars (because
they are affected by the coordinate transformation), nor can they be
tensor components of some kind (because they do not transform
linearly). As far as we know, no geometrical object that has yet been
defined in the mathematical literature can be used to describe the
$p_\mu$. Of course, all this discussion relies heavily on the use of
an ordinary spacetime manifold, which one might argue is not a
legitimate concept in the Planckian regime.  However, with no
spacetime manifold (and hence no notion of tangent vectors, tensors,
{\em etcetera\/}), the mathematical status of such objects as $p_\mu$
becomes even more mysterious.%

This point can be further clarified trying to rewrite equations
(\ref{msE}) and (\ref{msp}) in an explicitly covariant form. Since the
denominator $1-\pi_0/\kappa$ that appears in both equations
contains the pseudo-energy $-\pi_0$, there are only two ways in which
these equations can be interpreted, given that $\pi_0$ and $\pi_i$ are
the components of a one-form:%
\begin{enumerate}%
\item %
Suppose that equations (\ref{msE}) and (\ref{msp}) are valid in every
Lorentzian chart.  Then we can write%
\begin{equation}%
p_\mu=\frac{\pi_\mu}{1-\pi_\nu\,{\delta^\nu}_0/\kappa}\;,%
\label{ms-shit}%
\end{equation}%
where ${\delta^\mu}_\nu$ is the Kronecker symbol.  But by doing this
one introduces the chart-dependent structure ${\delta^\mu}_0$, which
would be regarded as meaningless in ordinary differential geometry.%
\item %
In contrast, suppose that equations (\ref{msE}) and (\ref{msp}) are
valid only in one particular class of Lorentzian coordinates.  Now we
can rewrite them in a covariant form as%
\begin{equation}%
p_\mu=\frac{\pi_\mu}{1-\pi_\nu\,u^\nu/\kappa}\;,%
\label{ms-cov}%
\end{equation}%
where $u^\mu$ is a four-vector that, in the preferred class of
coordinates, has components $u^0=1$ and $u^i=0$. But while this option
is perfectly sound from a mathematical point of view, the use of the
preferred vector $u^\mu$ unfortunately amounts to introducing a
preferred frame and an explicit breaking of Lorentz invariance, which
is in contrast with the whole spirit inspiring DSR
theories.\footnote{It is interesting to note that this case {\em is
    not equivalent} to the ansatz considered in most of the papers on
  quantum gravity phenomenology~\cite{QGph}. In these works energy and
  momentum are characterized by modified dispersion relations like
  (\ref{eq:disprel}), but they compose in the standard way (like
  $\pi_{\mu}$ does in our framework). In this sense they appear as
  hybrid models from the point of view of the situation envisaged by
  equation (\ref{ms-cov}).}
\end{enumerate}%
One way out of this dilemma (and in fact we suspect it is the only
mathematically sensible way out of this dilemma) is to reinterpret
equation (\ref{ms-cov}) without assuming that the four-velocity
$u^\mu$ is a preferred vector in spacetime.  Since the motivation for
the anomalous dispersion relation (\ref{mod-disp}) is ultimately of a
phenomenological character, one may interpret $E$ and $\mbox{\boldmath 
$p$}$ as the energy and three-momentum {\em measured\/} by a specific 
observer with four-velocity $u^\mu$.  Then, $E$ and the magnitude of the
three-momentum are true scalar quantities, representing the outcomes
of measurements performed by one particular specified observer, and so
are unaffected by coordinate changes.  But in which sense, then, can
one say that they ``transform nonlinearly under the action of the
Lorentz group?''%

\section{DSR as a new theory of measurement}%
\label{sec:measurement}%
\setcounter{equation}{0}%

The reformulation of $E$ and $\mbox{\boldmath $p$}$ in terms of an 
explicit observer-dependent four-velocity amounts, basically, to 
changing the theory of measurement in special relativity.  We now 
outline a modified theory of measurement in which DSRs can fit, and 
present a few speculations about the possible physical origin of the 
differences with respect to the ordinary theory.%

\subsection{``Real'' {\em versus\/} ``measured'' energy-momenta}%

Let us begin by recalling the important distinction between
coordinates (with no direct physical meaning, in general) and a {\em
  reference frame\/}, which is a field of tetrads
$\{e^\mu{}_\alpha\,|\,\alpha=0,1,2,3\}$ such that%
\begin{equation}%
\g_{\mu\nu}\,e^\mu{}_\alpha\,e^\nu{}_\beta=\eta_{\alpha\beta}\;,%
\label{gee}%
\end{equation}%
where now $\g_{\mu\nu}$ is the metric tensor,
$\eta_{\alpha\beta}=\eta^{\alpha\beta}=\mbox{diag}(-1,1,1,1)$, 
and $e^\mu{}_0$ is a future-directed vector. (Warning: The indices
$\mu,\nu,\ldots$ are standard tensor indices, associated with a choice
of coordinates, while the indices $\alpha$, $\beta$, $\ldots$ only
label different vectors in the tetrad, and have nothing to do with any
particular chart adopted.) Two reference frames $e^\mu{}_\alpha$ and
$\bar{e}^\mu{}_\alpha$ are related by a Lorentz matrix
${\Lambda_\alpha}^\beta$, so%
\begin{equation}%
\bar{e}^\mu{}_\alpha={\Lambda_\alpha}^\beta\,e^\mu{}_\beta\;.%
\label{lorentz}%
\end{equation}%

The use of a reference frame is crucial in order to extract, from the
abstract tensors of any relativistic theory, scalar quantities that
could be interpreted as measurement outcomes. In particular, in the
usual theory of measurement \cite{defelice}, if a particle has
four-momentum $\pi_\mu$, its energy and $i$-th component of
three-momentum, measured in the reference frame $\{e^\mu{}_\alpha\}$,
are given by the expressions:%
\begin{equation}%
\epsilon(\pi;e)=-\pi_\mu\,e^\mu{}_0\;;%
\label{E-standard}%
\end{equation}%
\begin{equation}%
\pi_i(\pi;e)=\pi_\mu\,e^\mu{}_i\;.%
\label{p-standard}%
\end{equation}%
Equations (\ref{E-standard}) and (\ref{p-standard}) can be
summarised into the single relation%
\begin{equation}%
\pi_\alpha(\pi;e)=\pi_\mu\,e^\mu{}_\alpha\;,%
\label{P-standard}%
\end{equation}%
under the identification $\pi_0\equiv-\epsilon$.%

It is important to realise that, while the $\pi_\mu$ are components of
a one-form in some chart, the $\pi_\alpha$ are {\em scalars\/} ---
\ie, they are a set of four chart-independent numbers.  However, they
depend on the reference frame adopted.  In a new frame
$\bar{e}^\mu{}_\alpha$, from the same one-form $\pi_\mu$ one obtains
four different scalars $\bar{\pi}_\alpha$, related to the $\pi_\alpha$
as%
\begin{equation}%
\bar{\pi}_\alpha=\pi_\mu\,\bar{e}^\mu{}_\alpha
=\pi_\mu\,{\Lambda_\alpha}^\beta\,e^\mu{}_\beta=
{\Lambda_\alpha}^\beta\,\pi_\beta\;,%
\label{linear}%
\end{equation}%
which is a usual linear Lorentz transformation.  In particular,
the $\bar{\pi}_\alpha$ are linear functions of the $\pi_\alpha$.%

In a DSR context we now suggest reformulating equation (\ref{F}) as%
\begin{equation}%
p_\alpha={\cal F}_\alpha(\pi_\mu\,e^\mu_0,
\pi_\mu\,e^\mu_1,\pi_\mu\,e^\mu_2,\pi_\mu\,e^\mu_3;\kappa)
= {\cal F}_\alpha(\pi_\mu\,e^\mu{}_\beta;\kappa)
\;,%
\label{F1}%
\end{equation}%
so that the ``physically measured'' energy and momentum, $E$ and
$p_i$, become nonlinear functions of both the underlying one-form
$\pi_\mu$ {\em and\/} the reference frame specified by the tetrad
$e^\mu{}_\alpha$.  In particular, equations (\ref{msE}) and
(\ref{msp}) proposed by Magueijo and Smolin \cite{ms} are to be
rewritten as%
\begin{equation}%
p_\alpha=\frac{\pi_\mu\,e^\mu{}_\alpha}{1 -\pi_\mu\,e^\mu{}_0/\kappa}\;,%
\label{ms1}%
\end{equation}%
with $E=-p_0$, as usual.  If the ${\cal F}_\alpha$ are nonlinear, then
upon performing a Lorentz transformation the $\bar{p}_\alpha$ are also
nonlinear functions of the $p_\alpha$.  This defines a mathematically
precise and physically consistent sense in which the theory is
simultaneously Lorentz-invariant (and covariant), while the physical
(measured) energy and momentum do not transform linearly under a
change of reference frame.%

In summary, our proposal is that the one-form $\pi_\mu$ be interpreted
as the ``real'' energy-momentum, and the four scalars $p_\alpha$ as
``measured'' energy-momenta.  The transformation from one to the other
additionally depends on the reference frame of the detector as encoded
in the tetrad $e^\mu{}_\alpha$. That is,%
\begin{equation}%
p_\alpha= {\cal F}_\alpha(\pi_\mu\,e^\mu{}_\beta;\kappa) \;,
\qquad \pi_\mu=
\g_{\mu\nu}\,e^\nu{}_\alpha\,\eta^{\alpha\beta}\,{\cal
G}_\beta(p_\gamma;\kappa)\;,%
\label{F1bb}%
\end{equation}%
where in the last equality we have used the completeness relation for
the tetrad,%
\begin{equation}%
\eta^{\alpha\beta}\,e^\mu{}_\alpha\,e^\nu{}_\beta=\g^{\mu\nu}\;.%
\end{equation}%
We further simplify this by defining%
\begin{equation}%
{\cal G}_\mu(p_\alpha;e;\kappa) :=
\g_{\mu\nu}\,
e^\nu{}_\alpha\,\eta^{\alpha\beta}\,
{\cal G}_\beta(p_\gamma;\kappa)\;,%
\label{F1bbb}%
\end{equation}%
so that%
\begin{equation}%
\pi_\mu ={\cal G}_\mu(p_\alpha;e;\kappa)\;.%
\label{F1bbbb}%
\end{equation}%
%

\subsection{Physical origin of the modifications}%

Up to this point, we have only inquired as to whether the formalism of
doubly special relativity can be made logically and mathematically
consistent, and we have refrained from asking physical questions.  In
this section we speculate about a possible physical basis for this new
interpretation of the DSRs.%

How should we understand the new theory of measurement expressed by
equation (\ref{F1})?  For the sake of clarity, let us focus on a
measurement of energy (similar considerations apply to measurements of
momentum components).  Setting the index $\alpha=0$ in equation
(\ref{F1}) we find that, in general, the measurement outcome for
energy (\ie, $E\equiv -p_0$) differs from the one predicted by
standard measurement theory (where we would obtain $\epsilon=-\pi_\mu
\; e^\mu{}_0$). We want to understand the origin of this discrepancy;
namely, how is it that a measurement does not reveal directly the
value $\epsilon$, given by equation (\ref{E-standard}), but the more
complicated expression given by equation (\ref{F1})?%

First of all, let us note that, in general, we can write%
\begin{equation}%
E=\epsilon+f(\epsilon;\kappa)\;,%
\label{exp}%
\end{equation}%
where $f$ is a function such that $f(\epsilon;+\infty)=0$.  The
discrepancy between $\epsilon$ and $E$ is thus due to the finiteness
of the DSR scale $\kappa$, which is usually taken to be the Planck
scale, since we assume the DSRs arise through quantum gravity effects.
In fact, based on the presumed existence of a smooth limit as gravity
is switched off, one might plausibly expect the dimensional parameter
$\kappa$ to lead to a relation such as%
\begin{equation}%
E = \epsilon \, \left[ 1 + \tilde f(\epsilon/\kappa) \right]\;,%
\end{equation}%
where $\tilde{f}(0)=0$.  This expression has the benefit of
reproducing the general form of the phenomenological models that have
been investigated in the literature.%

{From} these remarks, it seems plausible to identify the physical
origin of the discrepancy between the usual and the modified formulas
for the measured energy in the quantum gravitational effects that take
place whenever one performs a measurement.  If such effects were not
present, the measurement outcome for a particle with four-momentum
$\pi$ would be $\epsilon$, as usual.  However, they are universal and
non-screenable, so they always modify the measurement outcome into
$E$: This is why the measurement theory has to be revised.%

In connection to this point, it is worth mentioning that there exists
a consistent theme in the literature (see \cite{heisenberg} and
references therein), in which gravitational effects add to the
standard quantum uncertainty, producing modified Heisenberg relations
of the type%
\begin{equation}%
\Delta x\,\Delta p\stackrel{>}{\sim} \hbar\left(1+\lambda\,\Delta
p^2/\kappa^2\right)\;,%
\label{mod-heis}%
\end{equation}%
with $\lambda$ a numerical coefficient of order
one.\footnote{Analogous results follow from specific models for 
quantum gravity.}  Now, modified uncertainty relations can be traced
back, formally, to modified commutators.  And the DSR variables obey
modified commutation relations \cite{cortes}.  This appears to support
our claim about the phenomenological character of energy and momentum
used in DSR --- the variables $E$ and $\mbox{\boldmath $p$}$.  If one
were able to remove the additional uncertainty due to gravity, one
would end up with standard Heisenberg relations, and standard
commutation relations, for the variables $\epsilon$ and
$\mbox{\boldmath $\pi$}$.  However, whether a concrete proposal along
these lines is viable is of course a matter of debate, due to our
present lack of knowledge about the nature and effects of gravity in
the quantum regime.%

Note that, within this interpretation, the ``real'' energy-momentum is
$\pi$, while the $p_\alpha$ are only measurement outcomes.  However,
an effective theory formulated in terms of the $p_\alpha$ includes, in
general, a violation of Lorentz invariance.  This entails measurable
physical consequences.  Although quantum gravitational processes do
not affect $\epsilon$, they do affect $E$, hence \eg\ thresholds. More
dramatically, since the $\pi_\alpha = \pi_\mu \, e^\mu{}_\alpha$
transform under the ordinary linear Lorentz group, the symmetry
implies that it is these variables that will obey conservation of
four-momentum~\cite{jv}. It then need not be the case that the
measurement outcomes $p_\alpha$ satisfy conservation of energy and
momentum, with the possibility of Planck-suppressed violations of
these conservation laws now being a real concern.  Indeed, in this
framework it is logical to address questions about particle reactions
(\eg, thresholds) by imposing energy-momentum conservation on $\pi$
and than expressing the result in the measured variables $p_\alpha$
(see, \eg, reference \cite{Seth} for a concrete example of this
procedure).%

\section{Implications of the proposal}%
\label{sec:implications}%
\setcounter{equation}{0}%

Regarding DSR as a new theory of measurement in the way we suggested,
leads one to re-evaluate its physical consequences.  Here we sketch a
few implications of the newly proposed framework.%

\subsection{The saturation problem}%

Our working hypothesis (or more precisely, speculation) does not
provide an automatic resolution to the saturation problem of DSR.
However it is clear from the overall idea (that differences in the
measured energies and momenta are due to the quantum gravitational
interaction) that no simple prescription for the energy and momentum
composition of macroscopic bodies applies. In the case of large
numbers of particles, decoherence phenomena will take
place\footnote{This could even be caused by the gravitons themselves 
\cite{calucci}.}  and the measurement of the mass of a classical
object might not be affected by graviton exchange with the balance
used. In this sense, our proposal simply implies that it should be
impossible to find a coherent {\em quantum\/} system whose overall
mass is larger than the Planck mass.%

Indeed, we note that the most extensive Bose--Einstein condensates
experimentally created to date contain about $10^6$ atoms~\cite{bec},
corresponding to a mass of about $10^8$ GeV. If the DSRs in fact
represent the correct way of doing quantum gravity phenomenology, and
if our interpretation of the DSRs as a modified theory of measurement
is the correct one, then the ``saturation problem'' may be viewed
as predicting a maximum attainable mass for a Bose-Einstein condensate, 
of order one Planck mass, corresponding to about $10^{17}$ Rb atoms.  
This is a robust qualitative prediction of the DSR framework, which 
is in principle testable (though technically challenging). Furthermore, 
since in this framework the limitation alluded to above is actually a 
limitation on the maximum mass of a coherent quantum system we can 
(more boldly and more speculatively) also tie this back to Penrose's 
speculations on the gravitationally-induced collapse of the 
wave-function~\cite{Penrose}.  While one cannot, given the current 
state of knowledge, guarantee that this is the way the universe 
actually works, the new interpretation of the DSRs provides both a 
consistent logical framework, and a physical reason to suspect that 
such effects may be possible.%

A more precise way of putting this is to realise that a measurement of
the energy-momentum of some composite object depends not only on the
``true'' energy-momentum and on the observer's reference frame, but
also on many details of the internal structure of the composite
object, its interaction with the detector, and the internal
construction of the latter.  Let us collectively denote these extra
variables as $X$, and so write:%
\begin{equation}%
p_\alpha= {\cal F}_\alpha(\pi_\mu\,e^\mu{}_\beta;\kappa;X)\;;
\qquad %
\pi_\mu= {\cal G}_\mu(p_\alpha;e;\kappa;X)\;.%
\label{F1a}%
\end{equation}%
In particular, among the additional variables $X$ one can 
place:%
\begin{itemize}%
\item%
The total number $N$ of elementary particles in the body whose
``physical'' energy momentum is to be measured.  It is vitally
important to note that in the current context, where the nonlinear
transform ${\cal F}$ represents a phenomenological description of the
measurement process, an $N$-dependence of this type is much more
physically reasonable than in theories where ${\cal F}$ is assumed to
be ``fundamental''~\cite{jv}.%
\item%
The ``renormalization scale'' $\bar\mu$ typically used within the
particle physics community. This has the effect of giving a sensible
physical meaning to the ``running'' of ``measured'' energy-momentum
with ``renormalization scale''.%
\item%
The type of DSR (DSR1, DSR2, DSR3, ...?) appropriate to the particular
detector.  That is: As a by-product of this new interpretation, it is
now clear why the fundamental principles of DSR theories allow so many
apparently quite different DSR models --- at least three standard DSR
implementations are widespread in the literature. The multiplicity of
DSRs is simply a reflection of the fact that there are many different
classes of ``detectors'' one could think of building, all of 
which would be compatible with the basic framework of the DSR 
interpretation we advocate in this article.%
\end{itemize}%
For a shorthand that retains the key aspects of the physics, we can
discard other variables and simply write:%
\begin{equation}%
p_\alpha= {\cal
F}_\alpha(\pi_\mu\,e^\mu{}_\beta;\kappa;N,\bar\mu,DSR) \;; \qquad %
\pi_\mu= {\cal G}_\mu(p_\alpha;e;\kappa;N,\bar\mu,DSR) \;.%
\label{F1ab}%
\end{equation}%
%

\subsection{Conservation laws}%

Recall that in terms of the ``true'' energy-momenta $\pi$ the
transformation laws are simple, while in terms of the ``measured''
energy-momenta $p_\alpha$ the transformation laws are complicated.
This is telling us that it is the ``true'' energy-momenta $\pi$ that
are related to whatever underlying symmetries that via Noether's
theorem lead to conservation laws. This observation, in one form or
another, has led to almost universal acceptance in the literature of
the fact that conservation laws should be implemented in terms of the
``true'' energy-momenta $\pi$ --- the ``Judes--Visser variables''
of~\cite{jv}.%

Indeed in the ``modified measurement'' interpretation of the DSRs
advocated in this article it is clear that there is no physical need
for the ``measured'' energy-momenta to satisfy conservation laws ---
and this natural lack of conservation laws at high energies and
momenta is a quite generic feature of the DSRs.  Equally well, the
occurrence of non-standard dispersion relations is no longer
``unusual'' or ``peculiar'', but must instead be seen as quite natural
and in fact inevitable.%

\subsection{Einstein's equations}%

While in the highly interacting quantum gravity regime there will
certainly be drastic modifications to the standard Einstein equations,
there is observationally a wide range of distance and time scales in
the solar system and beyond over which standard general relativity
(and in particular the standard Einstein equations) works well.  In
this regime we can meaningfully ask whether the gravitational field
couples to the ``true'' energy-momenta $\pi$ or the ``measured''
energy-momenta $p_\alpha$?%

Since the solar system contains many macroscopic objects of
super-Planckian mass, the Einstein equations would seem to prefer to
couple to energy-momentum variables that do not suffer from any
saturation problem. For instance, if we try to couple the
gravitational field to ``measured'' energy-momenta $p_\alpha$, then
the resulting metric will depend not only on the source, but on the
four velocity (and in fact the entire tetrad) of the observing
apparatus (plus the particle content of the source, the resolution
[``renormalization scale''] of the observer, and the internal
structure of the detector).  Thus, in parallel with the fact that the
energy-momenta satisfies the relations%
\begin{equation}%
p_\alpha= {\cal
F}_\alpha(\pi_\mu\,e^\mu{}_\beta;\kappa;N,\bar\mu,DSR) \;, \qquad%
\pi_\mu= {\cal G}_\mu(p_\alpha;e;\kappa;N,\bar\mu,DSR) \;,%
\label{F1abc}%
\end{equation}%
we would now be {\em forced\/} to distinguish a ``true'' metric
$\gamma_{\mu\nu}$ from a ``measured'' metric ${\g}_{\mu\nu}$, with
relations of the form:%
\begin{equation}%
\g_{\mu\nu}= 
{\widetilde{\cal F}}_{\mu\nu}(\gamma_{\mu\nu};e;\kappa;N,
\bar\mu,DSR)\;; 
\qquad%
\gamma_{\mu\nu} = {\widetilde{\cal G}}_{\mu\nu}
(\g_{\mu\nu};
e;\kappa; N, \bar\mu,DSR)\;.%
\label{F1b3}%
\end{equation}%
But, in an application of {\em reductio ad absurdum\/}, this
``measured'' metric depends not on the apparatus that is measuring the
metric, but on the apparatus that is measuring the composite object
that is used as the source for the Einstein equations.%

The only way out of this is to apply the Einstein equations directly
to the ``true'' metric with the ``true'' variables as source.  One
could then {\em independently\/} introduce the notion of a
``measured'' metric as in equation (\ref{F1b3}), but now depending on
whatever apparatus is measuring the metric, and then interpret the
``measured metric'' ${\g}_{\mu\nu}$ as a ``running metric'' that
depends on the observer's motion and the resolution of his
(metric-measuring) apparatus.  Then, the ``measured'' metric need not
--- and in general will not --- satisfy the Einstein equations.  But
since in the DSR framework we know that deviations from standard
physics must be both Planck suppressed and macroscopically suppressed
we expect%
\begin{equation}%
\g_{\mu\nu} = \gamma_{\mu\nu} \, 
\left[ 1 + {\cal O}(\bar \mu/\kappa,N) \right]\;,%
\end{equation}%
so that any deviations from the metric expected on the basis of the
usual Einstein equations should also be greatly suppressed.%

To (hopefully) clarify the situation a little further: Suppose someone
tells you that at position $x$ she has ``measured'' the presence of a
particle with four-momentum $p_\alpha$. After making enquiries regarding
the structure of the particle detector, one would invert the nonlinear
transform ${\cal F}$ to determine the ``true'' four-momentum
$\pi_\mu$. This can then be inserted into the Einstein equation to
determine the ``true'' metric at some other point $y$. After making
enquiries regarding the structure of the metric detector placed at
$y$, one can apply the appropriate nonlinear transform
${\widetilde{\cal F}}$ (distinct from the previous one) to determine
the ``measured'' metric at $y$.%

In a similar vein one can now think of a parallel ``$\kappa$-deformed
phenomenology'' that would apply to all branches of physics --- for
instance there would be DSR-distorted electric and magnetic fields,
{\em etcetera\/}. While calculating the specific form of these DSR
distortions in any given situation would be quite horribly
complicated, the present interpretation has the great virtue of being
logically consistent and allowing us to ask physically meaningful
questions.%

\section{Conclusions}%
\label{sec:conclusions}%
\setcounter{equation}{0}%

The key point to be taken from the present article is that by viewing
the DSRs as a modified theory of measurement, we can provide a
mathematically precise, logically coherent, and physically non-trivial
interpretation for the DSRs.  The previous lack (apart from the
considerations of~\cite{girelli}) of any such coherent physical
interpretation has seriously hampered developments in the field. Key
features of the ``measurement'' interpretation of the DSRs are:%
\begin{itemize}%
\item%
There does not seem to be any pressing need to go to non-commuting
coordinates.  At least for the time being, ordinary differential
geometry based on Lorentzian manifolds seems quite sufficient as a
framework.%
\item%
Conservation laws, and the Einstein equations, seem to preferentially
couple to the ``true'' energy-momenta, which transform linearly under
the Lorentz group.%
\item%
``Measured'' energy and momenta do not only transform nonlinearly
  under the Lorentz group, but are now quite naturally seen to obey
  nonstandard dispersion relations, to not satisfy standard
  conservation laws, and to not directly act as sources for the
  Einstein equations.%
\item%
The DSRs are now to be viewed as phenomenological theories, that
depend on the measurement apparatus.  In a limited sense this may be
viewed as a ``demotion'', but in another sense this new point of view
now guarantees that the existence of DSR-like effects is both natural
and ubiquitous --- as is quantum gravity itself.%
\item%
With this new ``measurement'' interpretation there can no longer be
any doubt about the falsifiability of DSR effects and scientific
status of specific DSR theories, so there is a clear path to
experimentally testing the DSRs.  Without the interpretation we have
argued for in this article, or something closely related thereto, the
DSRs run the very real risk of amounting to physically empty
mathematical manipulations akin to the coordinate transformations of
general relativity.%
\end{itemize}%

In summary, we feel that the considerable confusion in the current
literature regarding the questions of consistency, triviality, and
physical acceptability of the DSRs is largely the result of
misinterpreting what the DSRs are trying to say. Viewed as a modified
theory of measurement, the DSRs make perfectly sensible statements
about empirical reality that can (at least in principle) be tested in
the usual scientific manner.%

\section*{Acknowledgements}%

The research of Matt Visser was supported by the Marsden fund
administered by the Royal Society of New Zealand.%

%
\end{document}